\documentclass[sigconf]{acmart} 

\AtBeginDocument{%
  \providecommand\BibTeX{{%
    \normalfont B\kern-0.5em{\scshape i\kern-0.25em b}\kern-0.8em\TeX}}}

\usepackage{amsfonts}
\usepackage{algorithmic}
\usepackage{textcomp}
\usepackage{listings}
\usepackage{makecell}
\usepackage{multirow}
\usepackage{svg}
\usepackage{xspace}
\usepackage{subfigure}
\usepackage{color, colortbl}
\usepackage{url}
\usepackage{pifont}

\newcommand{\circleone}{\ding{172}} 
\newcommand{\circletwo}{\ding{173}} 
\newcommand{\circlethree}{\ding{174}}

\DeclareUnicodeCharacter{2212}{-}
\definecolor{Gray}{gray}{0.9}
\definecolor{green}{RGB}{102,252,102}
\definecolor{ored}{RGB}{255,99,71}
\definecolor{orange}{RGB}{255,165,0}
\definecolor{lightgray}{RGB}{211,211,211}

\definecolor{lightgray}{gray}{0.93}
\definecolor{slightgray}{gray}{0.98}
\definecolor{darkgray}{gray}{0.77}
\definecolor{applegreen}{rgb}{0.55, 0.71, 0.0}
\definecolor{chromeyellow}{rgb}{1.0, 0.65, 0.0}
\definecolor{darkpastelgreen}{rgb}{0.01, 0.75, 0.24}

\newcommand{\tool}{{\textsc{MAdroid}}\xspace}

\newcommand{\chen}[1]{\textcolor{red}{\textbf{Chen}: #1}}

\copyrightyear{2026}
\acmYear{2026}
\setcopyright{rightsretained}
\acmConference[ICSE '26]{2026 IEEE/ACM 48th International Conference on Software Engineering}{April 12--18, 2026}{Rio de Janeiro, Brazil}
\acmBooktitle{2026 IEEE/ACM 48th International Conference on Software Engineering (ICSE '26), April 12--18, 2026, Rio de Janeiro, Brazil}
\acmPrice{}
\acmDOI{10.1145/3744916.3764535}
\acmISBN{979-8-4007-2025-3/26/04}

\begin{document}

\title[Multi-Agent LLMs for Multi-User Feature Testing]{Breaking Single-Tester Limits: Multi-Agent LLMs for Multi-User Feature Testing}

\author{Sidong Feng}
\affiliation{%
  \institution{Monash University}
  \city{Melbourne}
  \country{Australia}}
\email{sidong.feng@monash.edu}

\author{Changhao Du}
\affiliation{%
  \institution{Jilin University}
  \city{Changchun}
  \country{China}}
\email{chdu22@mails.jlu.edu.cn}

\author{Huaxiao Liu}
\affiliation{%
  \institution{Jilin University}
  \city{Changchun}
  \country{China}}
\email{liuhuaxiao@mails.jlu.edu.cn}

\author{Qingnan Wang}
\affiliation{%
  \institution{Jilin University}
  \city{Changchun}
  \country{China}}
\email{wangqn23@mails.jlu.edu.cn}

\author{Zhengwei Lv}
\affiliation{%
  \institution{Bytedance}
  \city{Beijing}
  \country{China}}
\email{lvzhengwei.m@bytedance.com}

\author{Mengfei Wang}
\affiliation{%
  \institution{Bytedance}
  \city{Beijing}
  \country{China}}
\email{wangmengfei.pete@bytedance.com}

\author{Chunyang Chen}
\affiliation{%
  \institution{Technical University of Munich \\ \& Monash University}
  \city{Heilbronn}
  \country{Germany}}
\email{chun-yang.chen@tum.de}

\renewcommand{\shortauthors}{Sidong Feng, et al.}

\begin{abstract}
The growing dependence on mobile phones and their apps has made multi-user interactive features—like chat calls, live streaming, and video conferencing—indispensable for bridging the gaps in social connectivity caused by physical and situational barriers. However, automating these interactive features for testing is fraught with challenges, owing to their inherent need for timely, dynamic, and collaborative user interactions, which current automated testing methods inadequately address. Inspired by the concept of agents designed to autonomously and collaboratively tackle problems, we propose \tool, a novel multi-agent approach powered by the Large Language Models (LLMs) to automate the multi-user interactive task for app feature testing. Specifically, \tool employs two functional types of multi-agents: user agents (Operator) and supervisor agents (Coordinator and Observer). Each agent takes a specific role: the Coordinator directs the interactive task; the Operator mimics user interactions on the device; and the Observer monitors and reviews the task automation process. Our evaluation, which included 41 multi-user interactive tasks, demonstrates the effectiveness of our approach, achieving 82.9\% of the tasks with 96.8\% action similarity, outperforming the ablation studies and state-of-the-art baselines. Additionally, a preliminary investigation underscores \tool's practicality by helping identify 11 multi-user interactive bugs during regression app testing, confirming its potential value in real-world software development contexts.
\end{abstract}



\maketitle

\section{Introduction}

\begin{figure}
	\centering
	\includegraphics[width = 0.95\linewidth]{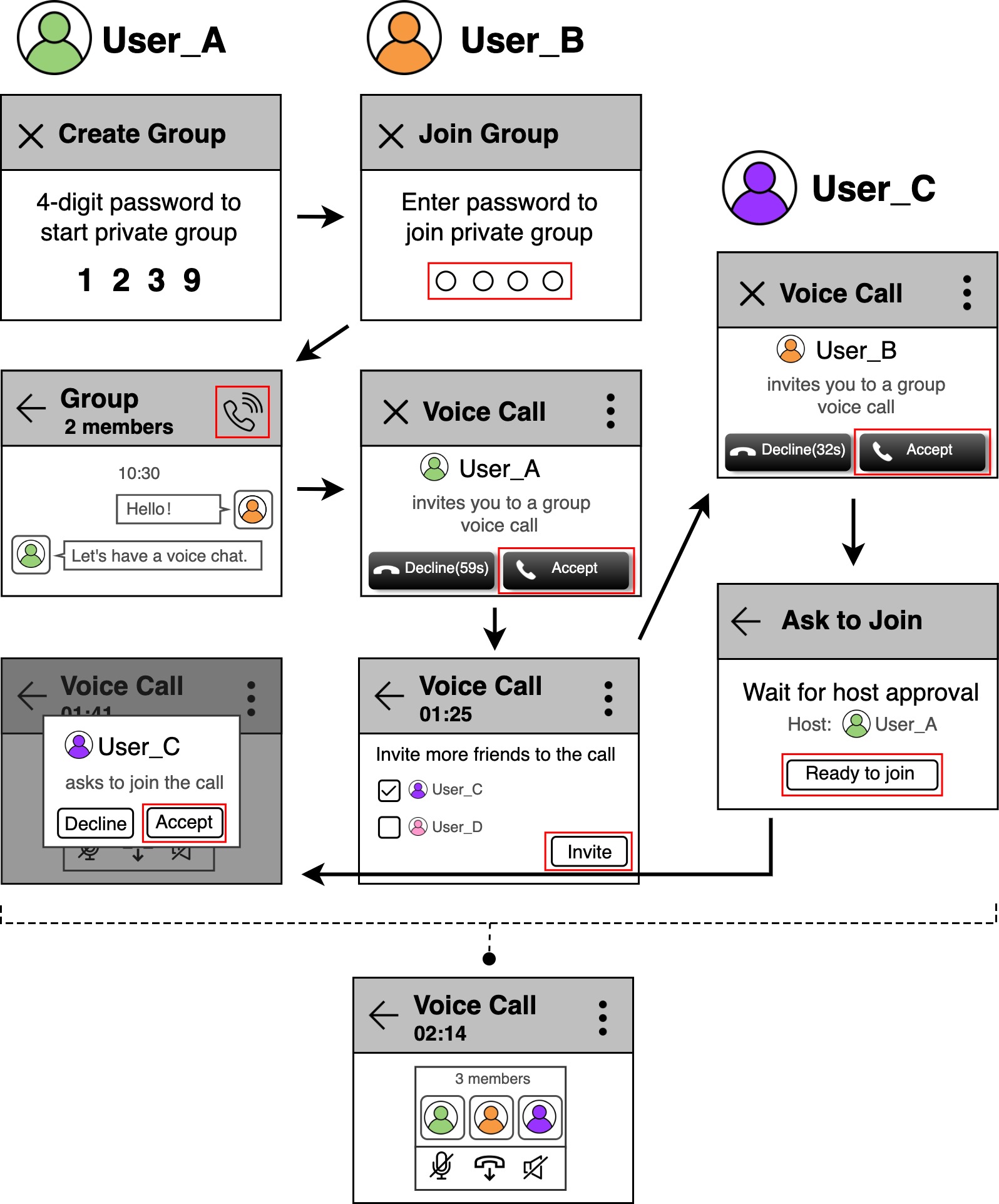}
	\caption{Illustration of a group voice call. }
	\label{fig:call}
\end{figure}

Given millions of applications (apps) available from app stores like the Google Play Store~\cite{web:google} and the Apple App Store~\cite{web:apple}, mobile phones and apps now have become indispensable for our daily lives in accessing the world.
Among the various features of mobile apps, multi-user interactive capabilities have become an essential element, particularly in the realm of social networking. 
The incorporation of interactive features such as voice calls, live streaming, and video conferencing, has transformed the way users engage with each other.
With social media apps being a notable example, they offer virtual communication alternatives when face-to-face interactions are impeded by distance, the COVID-19 pandemic, or other barriers to accessibility.
For instance, as shown in Figure~\ref{fig:call}, a user ($User_{A}$) can start a private conversation by generating a unique access code, enabling others ($User_{B}$) to join the chat using this code to a real-time communication session.
Additionally, they can extend invitations to mutual friends ($User_{C}$) to participate in collaborative group chats.
This straightforward multi-user interactive feature facilitates immediate communication, collaboration, and social connection, bridging physical divides.

Despite their importance, multi-user interactive features are often overlooked during automated app testing~\cite{chen2023unveiling,feng2024enabling,feng2024mud}.
This oversight stems from two key challenges.
First, triggering these features requires multiple users to interact within the app in a coordinated manner, as these features typically involve collaborative actions that cannot be activated by a single user operating in isolation.
For example, a group voice call feature in Figure~\ref{fig:call}, requires three users, e.g., one user to create the call and others to join, each from their separate device. 
Second, the interactions between multiple users are inherently sequential.
The correct sequence of interactions, such as starting and accepting the call, must be executed in a rigid order.
Any discrepancy in the number of users or mistakes in the execution order of these interactions will lead to the failure of the entire testing process.
While extensive automated app testing tools~\cite{web:monkey,li2017droidbot,li2019humanoid,liu2023make} have focused on enhancing algorithms to improve testing coverage, none have sufficiently addressed the challenge of dynamic and collaborative interactions across multiple users, perpetuating the inherent limitation of low testing coverage for multi-user interactive features.

Some research~\cite{kumaresen2020agent,ramirez2021agent,ahlgren2021behavioural,ahlgren2020wes} has introduced the concept of multi-agent systems for software testing, where each agent is an entity designed to autonomously perceive and interact with its environment, and solve problems cooperatively. 
However, these systems tend to rely on hard-coded rules specific to individual apps, limiting their generalizability across different software.
In the last few years, AI multi-agents have shown human behavior patterns~\cite{wang2023survey}, and even show superhuman performance in games like AlphaStar for Starcraft~\cite{vinyals2019grandmaster}, and OpenAI Five for Dota 2~\cite{berner2019dota}. 
The emerging Large Language Models (LLMs) such as GPT~\cite{brown2020language},  Claude~\cite{web:claude}, and Mistral~\cite{web:mistral} make it even harder by providing a backbone of autonomous agents that can interact with each other like humans~\cite{park2023generative}. 
Inspired by the role-playing capability of LLMs, we propose a novel multi-agent system powered by the LLMs to autonomously conduct multi-user interactive feature testing, with agents assuming distinct, human-like roles for covering multi-user interactive features.

In this paper, we introduce \tool, a novel approach designed to autonomously interact with mobile apps, facilitating the automation of multi-user interactive features by utilizing multi-agent LLMs.
Based on the feature testing paradigm studied in previous research~\cite{feng2024enabling,feng2025agent}, we define high-level natural language descriptions of multi-user interactive features as test inputs.
Our approach then distinctly categorizes agents into two functional types: user agents (i.e., \textit{Operator}) and supervisor agents (i.e., \textit{Coordinator}, \textit{Observer}). 
User agents are designed to mimic end-users, simulating multi-user interactions within the app environment, while supervisor agents with distinct roles aim to oversee the execution of tasks.
In detail, given a description of a multi-user interactive feature, a \textit{Coordinator} is to interpret, organize, and assign the task. 
This includes deciding the number of necessary users, distributing task segments to the specific user agents, and determining the order in which these task segments should be carried out.
Subsequently, multiple \textit{Operator}s are responsible for navigating through the GUI screens to perform the tasks assigned to them.
However, the Operator's view is confined to its own GUI screen, unaware of the broader context, which could introduce a potential bias. 
To address this, we incorporate an \textit{Observer}, that has access to a higher level of insight, including information from the task coordination and a record of actions and screens carried out by different Operators, to monitor and review the overall progression of the task.
Through this multi-agent structure, \tool bridges the gap between automated task execution and the nuanced dynamics of multi-user interaction.

We conducted a comprehensive evaluation to assess the effectiveness of our \tool in testing multi-user interactive features. 
As the first work in this direction, we manually constructed a dataset comprising 41 descriptions of multi-user interactive feature tasks and their corresponding ground-truth GUI traces.
Results demonstrate that our \tool achieves a success rate of 82.9\% and 96.8\% of its actions were the same as the ground truth, significantly outperforming three ablation studies and eight state-of-the-art baselines.
We also examined the usefulness of our \tool in real-world regression testing scenarios by automating 37 multi-user interactive task descriptions over three app versions.
Throughout the process, our approach helps identify 11 interactive bugs, indicating initial evidence of \tool in enhancing the testing process.
The contributions of this paper are as follows:
\begin{itemize}
    \item This is the first work to define the problem of multi-user interactive features within the software, an aspect that has previously been overlooked in automated testing.
    \item We propose a novel approach, \tool, that harnesses a multi-agent LLMs framework for the autonomous automation of multi-user interactive tasks. 
    The source code, dataset, and experimental results are in appendix\footnote{https://github.com/sidongfeng/MAdroid}.
    \item Comprehensive experiments, including the effectiveness and usefulness evaluation of \tool and its detailed qualitative analysis for interactive task automation.
\end{itemize}

\section{Definition \& Background}
Unlike the general features, the app's multi-user interactive feature encompasses collaborative interactions that are responsive to inputs from multiple users. 
In this section, we first define the problem of multi-user interactive features. 
Then, we briefly discuss the concept of multi-agent that we adopt in the design of our approach.



\subsection{Definition of Multi-User Interactive Feature}
\label{sec:definition}


\begin{center}
\vspace{3pt}
\fbox{%
  \parbox{0.95\linewidth}{%
  \vspace{3pt}
    \textbf{Definition}: \textit{Multi-user interactive feature is a component of software that enables \textsuperscript{\circleone}\underline{multiple users} to \textsuperscript{\circletwo}\underline{interact with each other} in \textsuperscript{\circlethree}\underline{nearly real-time} within a software.}%
\vspace{3pt}
  }%
}
\vspace{3pt}
\end{center}

The definition of multi-user interactive features is dissected into three fundamental elements.
\circleone The feature enables user-to-user communication through the software.
It can be a two-way or multi-way interaction between end-users.
Note that interactions between a user and the software or the administrator are not categorized under multi-user interactive features. 
An example of this would be activity organization, where an end-user books a reservation and the admin confirms it. 
Such interactions are predominantly administrative and do not embody the peer-to-peer collaborative or communicative intent that is the hallmark of multi-user interactive features.
\circletwo A critical aspect of the interactive feature is its fundamental interactivity.
This characteristic is marked by the presence of interactive cues including notifications or pop-up messages, that differ significantly from collaborative features like text editing.
These cues play a crucial role in ensuring users are aware of others' requests and encouraging them to act or respond.
\circlethree The interactive feature pertains to a timely nature.
Interactions should occur with minimal delay, though not necessarily instantaneously, to maintain a fluid information exchange and keep users actively engaged. 
For instance, the call will automatically expire after 59 seconds in Figure~\ref{fig:call}, which the current automated testing frameworks fail to proceed to further communication activities, potentially impeding testing coverage.



\subsection{Background of Multi-agents}
Agents~\cite{wooldridge1997agent} are software entities designed to perceive and interact with their environment, capable of performing autonomous actions.
Depending on the flexibility of such actions, agents possess the ability to independently initiate tasks and set their own objectives.
Nonetheless, when addressing complex and collaborative problem-solving scenarios, relying solely on a single agent is often insufficient. 
Such scenarios necessitate the adoption of a multi-agent framework, a coalition of multiple model-driven agents that work collaboratively. 
Extending beyond the output of single-agent to provide solutions, the multi-agent framework deploys a team of agents, each bringing specialized knowledge to collectively address sophisticated challenges.
In these setups, several autonomous agents collaborate in activities akin to human teamwork, such as planning, discussion, and decision-making in problem-solving endeavors. 
Essentially, this approach harnesses the communicative strengths of agents, utilizing their generation and response capabilities for interaction.
In our research of testing multi-user interactive features within the software, where interactions between multiple users involve dynamic exchanges of input and feedback, we have embraced the multi-agent paradigm to enhance the collaboration of agents, making it a valuable asset for interactive features.

\section{Approach}

\begin{figure}
	\centering
	\includegraphics[width = 0.8\linewidth]{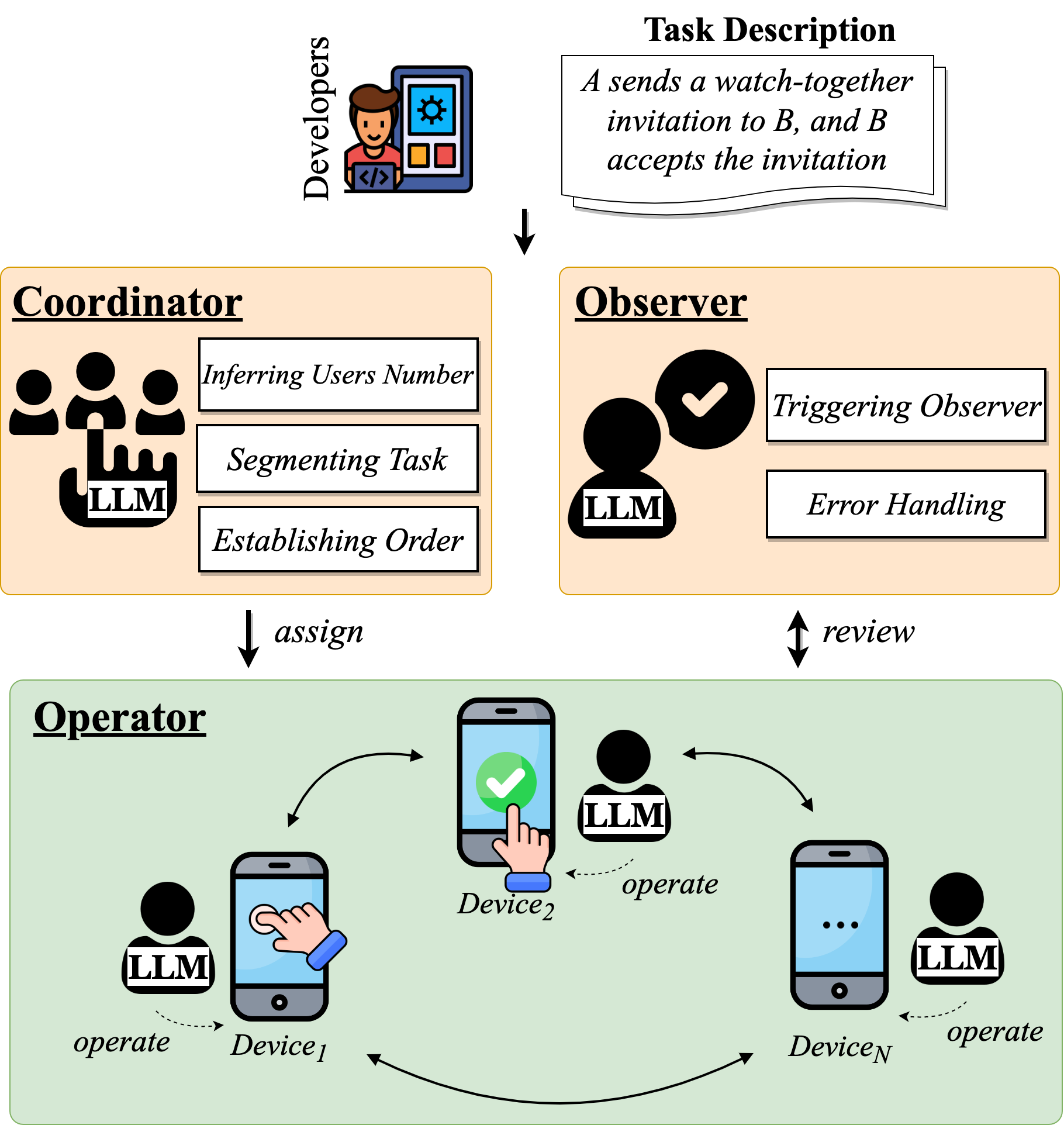}
	\caption{The overview of \tool.}
	\label{fig:overview}
\end{figure}

\tool is designed with a multi-agent-based architecture, including user-oriented agents (\textit{Operator} in Section~\ref{sec:phase2}) and supervisor-oriented agents (\textit{Coordinator} in Section~\ref{sec:phase1} and \textit{Observer} in Section~\ref{sec:phase3}).
The overview of our approach is shown in Figure~\ref{fig:overview}.
Specifically, the Coordinator is to plan the interactive task, the Operator interacts on devices, and the Observer reviews the task progress.
This structure mirrors the analogous human roles typically involved in multi-user interactive feature testing: the task director, the interacting users, and the task reviewer, respectively. 


\subsection{Coordinator}
\label{sec:phase1}
\begin{figure}
	\centering
	\includegraphics[width = 0.96\linewidth]{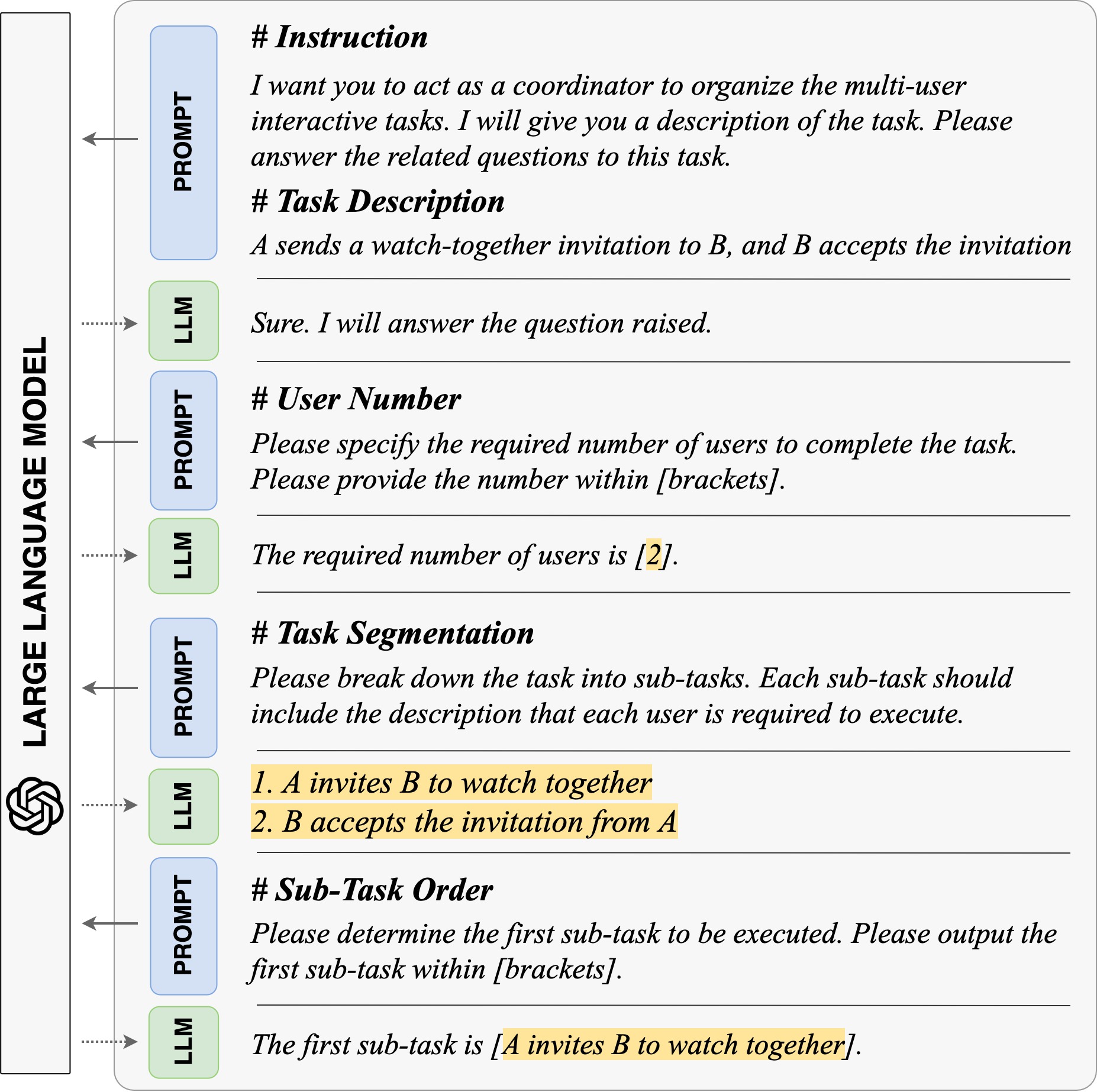}
	\caption{The example of prompting Coordinator. }
	\label{fig:coordinator}
\end{figure}

The initial stage of our approach is to utilize the Large Language Models (LLMs) to understand, analyze, organize, and strategize for prospective events derived from a high-level interactive task description ($task$).
An example of the prompts given to LLMs as Coordinator is detailed in Figure~\ref{fig:coordinator}.
In detail, it first commences by interpreting the instructions as shown in Figure~\ref{fig:coordinator}-A, which acts as a preliminary framework guiding the LLMs within the task coordination workflow.
Subsequently, the agent assumes the role of autonomous planner, which involves figuring out the number of required users, partitioning the task into each user, and determining the task order. 
Note that our task planning is executed step by step, aligning with the chain-of-thought paradigm~\cite{feng2023prompting}.

\subsubsection{Inferring user number}
Our initial prompt aims to engage LLMs in determining the required number of interactive users needed for collaboration on the interactive feature.
This is accomplished by presenting the LLMs with a prompt to specify the exact number of users essential for the successful completion of the task, in Figure~\ref{fig:coordinator}.


\subsubsection{Segmenting task}
Upon determining the requisite number of users for the task, we then need to segment the $task$ into smaller $sub$-$tasks$ that can be distributed across these interactive users. 
This involves a potential plan of the task to identify discrete components that can be independently executed by different devices. 
For example, consider the task of \textit{``A sends a watch-together invitation to B, and B accepts the invitation''} in Figure~\ref{fig:coordinator}.
It will be broken down into sub-tasks, including \textit{``A invites B to watch together''} for $user_A$ and \textit{``B accepts the invitation from A''} for $user_B$.
The detail of the prompt is shown in Figure~\ref{fig:coordinator}.

\subsubsection{Establishing sub-task order}
When dealing with interactions between users, the sequence in which they undertake their respective sub-tasks is crucial. 
Taking the watch-together scenario as an example, the correct sequence would have $user_A$ initiating the invitation; this action must logically come before $user_B$ can accept the invitation. 
To facilitate this process, we use prompts for the LLMs to identify the initial sub-task to be executed, as outlined in Figure~\ref{fig:coordinator}.
Note that our focus is on determining the first sub-task, i.e., the initial event carried out by a user, rather than mapping out the entire sequence of sub-tasks. 
This is because the interaction between users can introduce complexities. 
For instance, after $user_B$ accepts the invitation, $user_A$ may need to provide additional permissions for the watch together connection to be fully established, which is a subsequent interaction that can arise during the execution of sub-tasks.

\subsection{Operator}
\label{sec:phase2}
Once the task coordination has been determined, we proceed by selecting the devices from the device farm, assigning them their sub-tasks, and establishing their initiation sequence. 
Subsequently, we deploy an individual Operator agent onto each device to navigate the GUI screen and fulfill the designated task. 
However, employing LLMs to operate the devices presents two primary challenges:
First, LLMs are not inherently equipped for GUI interaction, which may result in a lack of operational knowledge pertaining to the devices.
To address this, we supply the LLMs with a list of actions and their corresponding primitives to facilitate interaction with the GUI screen.
Second, GUIs are typically rich in spatial information, embodying a structural complexity that LLMs might not readily comprehend from visual GUI screenshots~\cite{yang2023dawn}. 
To overcome this, we introduce a heuristic approach that converts the GUIs into domain-specific representations, which are more accessible to the LLMs' understanding.

\subsubsection{Action space}
Building on prior work on GUI automation~\cite{feng2023prompting,ran2024guardian}, we define the core actions for multi-user interaction, as presented in Table~\ref{tab:action}.
These include three standard actions, including tap, input, and back, as well as two task-specific actions, including switch\_user, and end\_task.
While additional customized actions like pinch and multi-finger gestures exist, they are less prevalent.
For brevity, we focus on the commonly-used actions in this paper.

The representation of each action is contingent on its specific context, necessitating distinct primitives to encapsulate the involved entities. 
For instance, the ``tap'' action mandates a target element within the GUI, like a button. 
Therefore, we express this as [tap] [element]. 
When it comes to the ``input'' action, it entails inputting a designated value into a field, and we structure this as [input] [element] [value]. 
Additionally, we accommodate a system action that does not stem directly from the current GUI screen: ``back'', which allows for navigation to a previously visited screen.

We extend our support to actions specifically designed for interactions. 
Upon approaching the potential completion of a sub-task, one user sends a signal to switch control to another user to continue the work in sequence. 
In addition, this process may also involve the transfer of additional communication details such as a ``joining code''. 
To represent this interactive communication action, we employ the notation [switch] [user] [message]. 
Additionally, we have defined an [end\_task] action, which serves as an indicator for models to identify the completion of the task.

\renewcommand{\arraystretch}{1.05}
\begin{table}
        \scriptsize
	\centering
	\caption{Action space for Operator.}
	\label{tab:action}
	\begin{tabular}{l|l|p{4cm}} 
		\hline
		\rowcolor{darkgray} \bf{Action}  & \bf{Primitive} & \bf{Description} \\
		\hline
		  tap & [tap] [element] & click on the element on the GUI screen. \\
            \hline
            input & [input] [element] [value] & enter the value into the GUI element field. \\
            \hline
            back & [back] & return to the preceding GUI screen. \\
            \hline
            switch\_user & [switch] [user] [message] & switch to a different user with a message. \\
            \hline
            end\_task & [end\_task] & finish the task. \\
            \hline
	\end{tabular}
\end{table}

\subsubsection{GUI screen representation}
To guide LLMs in navigating GUIs, it is essential to dynamically present the contextual information of the current GUI screen.
For this purpose, we represent a mobile GUI's structure in XML view hierarchy format.
These XML representations map out the hierarchical relationships among GUI elements and include various attributes, like class names, text, descriptions, and clickability, that convey the functionality and visual characteristics of the GUI elements. 
However, these attributes can sometimes contain extraneous details, such as text color, shadow color, and input method editor (IME) options, which may complicate the LLMs' ability to interpret the screen.

To address this, we first simplify the GUI elements, by removing non-essential attributes to concentrate on those that are critical for comprehension of the GUI.
In order to ascertain which attributes are indispensable, we conduct an in-depth review of the Android documentation~\cite{web:accessibility} and previous research~\cite{feng2023prompting,wang2023enabling,liu2023fill}, pinpointing GUI attributes that are semantically significant for interaction. 
Consequently, we opt to use a subset of properties from the element.


\begin{itemize}
    \item \textit{resource\_id}: describes the unique resource id of the element, depicting the referenced resource.
    \item \textit{class}: describes the native UI element type such as TextView and Button.
    \item \textit{clickable}: describes the interactivity of the element.
    \item \textit{text}: describes the text of the element.
    \item \textit{content\_desc}: conveys the content description of the visual element such as ImageView.
\end{itemize}

Next, we refine the screen layout to a cleaner depiction for LLMs understanding, by discarding the empty layout container.
For this purpose, we employ a depth-first search algorithm to navigate through the view hierarchy tree and identify nested layouts. 
We iterate over each node, beginning with the root, and eliminate any layouts that consist of a single node, subsequently proceeding to its child node. 

\begin{figure}
	\centering
	\includegraphics[width = 0.96\linewidth]{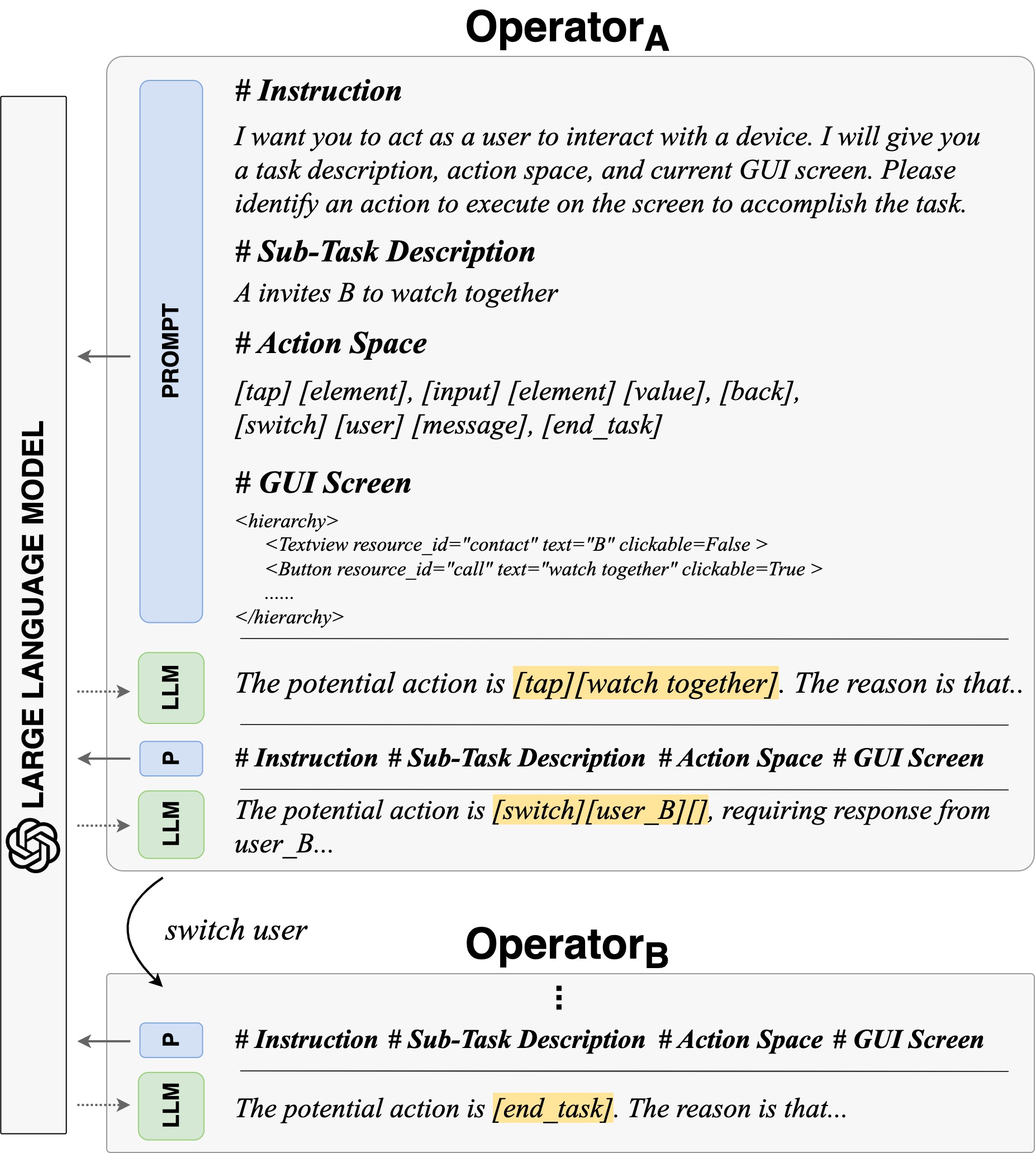}
	\caption{The example of prompting Operator.}
	\label{fig:operator}
\end{figure}

\subsubsection{Prompting LLMs}
Considering the allocated $sub$-$task$, the range of possible actions, and the current GUI screen, we prompt the LLMs to infer a single viable action to execute on the screen, thereby advancing one step toward the completion of the $task$. 
An illustrative example of such prompting is provided in Figure~\ref{fig:operator}. 
Due to the robustness of LLMs, the prompt sentence does not need to adhere strictly to grammar rules~\cite{brown2020language}.
For instance, given the sub-task to send an invitation to watch together, the output of the action to be performed on the current GUI screen would be [tap] [watch together].
The Operator then executes the suggested actions on the device by using the Android Debug Bridge (ADB)~\cite{web:adb}.
This cycle of prompting LLMs for a plausible action is repeated until the LLMs recognize the need to interact with a different user, at which point infer a transition command to another Operator, i.e., [switch] in Figure~\ref{fig:operator}.
When an Operator concludes that the $sub$-$task$ has been fully executed without the need for cross-user interaction, it generates a signal indicating task completion (i.e., [end\_task]).


\subsection{Observer}
\label{sec:phase3}
Integrating LLMs for achieving interactive actions does not guarantee flawless inference of the desired outcomes. 
This limitation stems from two prevalent challenges.
First, the Operator primarily processes lower-level information, focusing on the current GUI understanding for immediate action inference.
Whereas, access to higher-level information, such as the sequence of GUIs across various users, could enhance task automation.
Second, a single LLM agent may be susceptible to hallucinations~\cite{huang2023survey}, meaning they might generate fabricated actions, presenting them as factual steps toward task completion.
Any error in any step may fail the whole automation.
To mitigate these issues, similar to how professional developers carry out code reviews, we introduce an additional LLM agent acting as an external observer. 
This agent periodically provides feedback based on a higher level of information.

\begin{figure}
	\centering
	\includegraphics[width = 0.96\linewidth]{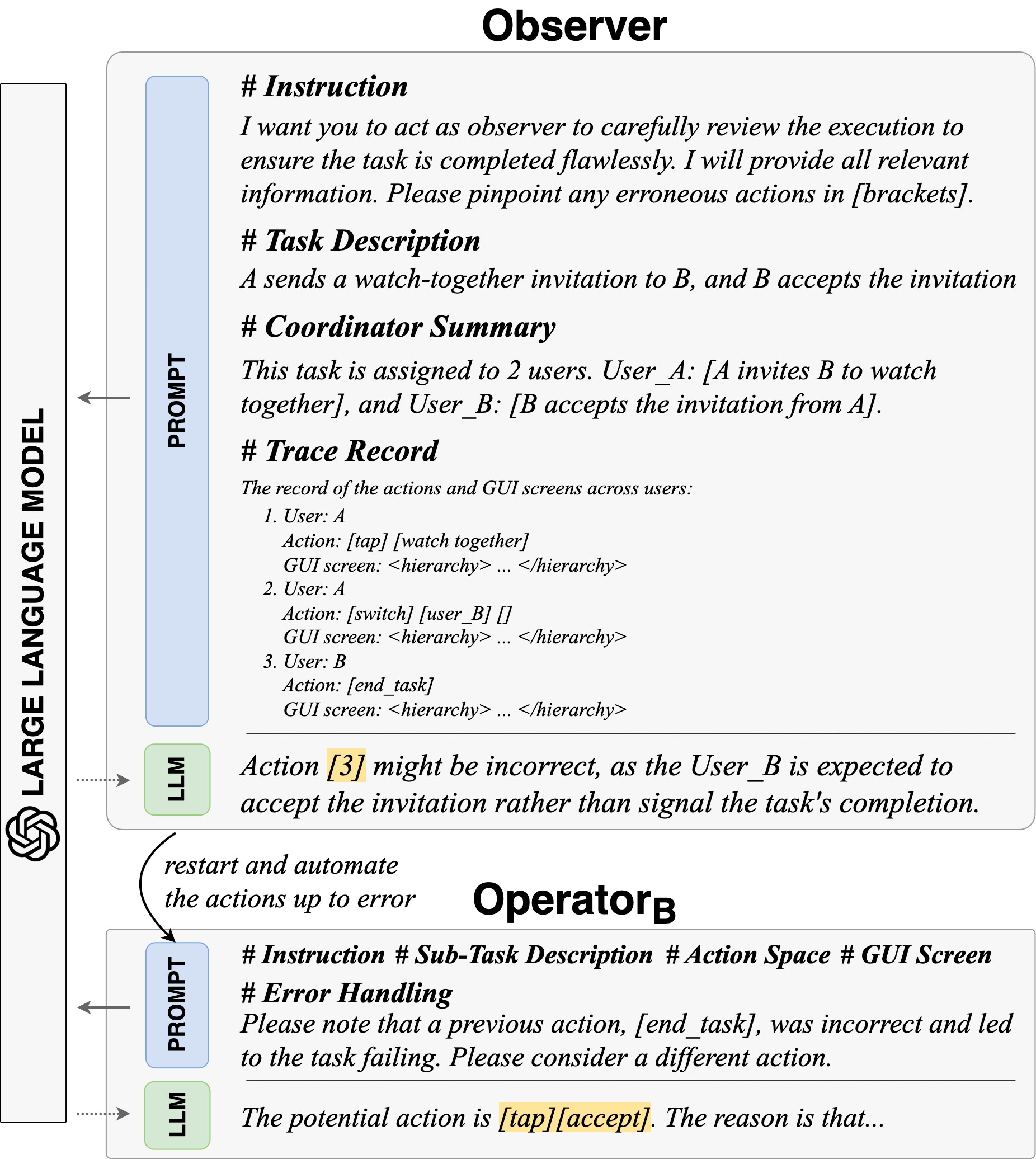}
	\caption{The example of prompting Observer.}
	\label{fig:observer}
\end{figure}

\subsubsection{Triggering Observer}
Determining the optimal activation for the Observer to review and provide feedback on the actions taken by the Operator is not trivial.
Reviewing each action is inefficient, while activating the Observer only at the end of the task to verify the correctness of actions is ineffective.
For instance, an error made halfway through could set the course on an incorrect trace, with the flaw only being detected after the task is complete. 
Moreover, such a misstep may create a repetitive cycle that potentially traps the task in an infinite loop. 
Hence, based on a small-scale pilot study, we have established a threshold where the Observer is triggered after every two actions,  striking a balance between effectiveness and efficiency.
We also activate the Observer after the task completion to confirm its success.

An illustrative example of prompting Observer is provided in Figure~\ref{fig:observer}. 
Specifically, we first instruct the agent with the role of detecting any potential erroneous actions and clarify the overall objective of the interactive task. 
We then present a summary of the task coordination from the Coordinator (Section~\ref{sec:phase1}), utilizing a customized natural language template.
Subsequently, we document the full sequence of actions executed across various users, which includes the users operated, the actions taken, and the GUI screens presented.
However, compiling such a full sequence may lead to a lengthy text for the LLMs to process, particularly toward the end of the task when the action execution sequence becomes extensive.
To address this issue, we truncate the record by removing the earliest actions performed on the screen.
Note that this record is formatted as a list to facilitate the identification of specific error actions by LLMs.

\subsubsection{Error handling}
Once an error is detected, we need to revert to the last known correct state and provide guidance to the specific Operator agent responsible for the erroneous action, directing them toward a correct course of action.
While reversing actions to a previous state might seem straightforward to achieve by the system backward, it fails for interactive actions, such as reversing a declined invitation.
Consequently, our approach tends to restart the task and automate the progression of actions up to the point just before the error occurred.
Then, to alert the specific Operator agent to the error, we append a sentence to the end of the prompt as shown in Figure~\ref{fig:observer}: ``\textit{Please note that a previous action, [previous action], was incorrect and led to the task failing. Please consider choosing a different action.}''
This serves to highlight the error and encourage the Operator to make a more accurate action.

\subsection{Implementation}
Our \tool has been developed as a fully automated tool for achieving multi-user interactive features within the apps given a task description. 
For LLMs, we use the state-of-the-art GPT-4 model from OpenAI~\cite{web:chatgpt} as our foundational model and assistant~\cite{web:assistant} to keep track of context.
It's important to note that we assign separate LLM instances to fulfill the roles of Coordinator (Section~\ref{sec:phase1}), each Operator (Section~\ref{sec:phase2}), and Observer (Section~\ref{sec:phase3}).
This is achieved by each running on different threads, in order to prevent any role confusion among the agents.
To automatically parse the LLMs' output, we provide structured formatting instructions that allow for straightforward interpretation using regular expressions (i.e., []). 
For the automation execution on devices, we employ Genymotion~\cite{web:genymotion} to allocate, initiate, and manage virtual Android devices.
We use Android UIAutomator~\cite{web:uiautomator} for extracting the GUI XML view hierarchy, and Android Debug Bridge (ADB)~\cite{web:adb} to carry out the actions to devices.
\section{Evaluation}
\label{sec:evaluation}
In this section, we describe the procedure we used to evaluate \tool in terms of its performance.

\begin{itemize}
    \item \textbf{RQ1:} How effective is our approach in automating multi-user interactive feature tasks?
    \item \textbf{RQ2:} How does our approach compare to the state-of-the-art methods?
    \item \textbf{RQ3:} How useful is our approach in real-world multi-user interactive feature testing?
\end{itemize}

For \textbf{RQ1}, we present the general performance of our approach in automating multi-user interactive tasks.
Moreover, we assess the impact of individual components within our approach by conducting an ablation study.
For \textbf{RQ2}, we carry out experiments to check the effectiveness of our approach against eight state-of-the-art baselines. 
For \textbf{RQ3}, we evaluate the usefulness of our approach to detect interactive bugs within real-world development environments.

 \renewcommand{\arraystretch}{0.95}
\begin{table*}
    \footnotesize
    \tabcolsep=0.1cm
	\centering
	\caption{Detailed results for 20 multi-user interactive tasks, and the actions are displayed in a format of \#successfully-executed/\#ground-truth. }
	\label{tab:rq1_performance_detail}
	\begin{tabular}{l|p{5.3cm}||c|c|c|c|c|c|c|c|c|c|c|c} 
	    \hline
	   \multirow{2}{*}{\bf{Task}} & \multirow{2}{*}{\bf{Description}} & \multicolumn{3}{c|}{\bf{Operator only}} & \multicolumn{3}{c|}{\bf{Operator+Coordinator}} & \multicolumn{3}{c|}{\bf{Operator+Observer}} & \multicolumn{3}{c}{\bf{\tool}}\\
	    \cline{3-14}
            & & $user_A$ & $user_B$ & $user_C$ & $user_A$ & $user_B$ & $user_C$ & $user_A$ & $user_B$ & $user_C$ & $user_A$ & $user_B$ & $user_C$ \\
            \hline
             
             \rowcolor{lightgray} Tiktok-1 & User$_{A}$ sends an interactive card to User$_{B}$, and User$_{B}$ activates the card & 3/3 & 1/1 & - & 3/3 & 1/1 & -  & 3/3 & 1/1 & - & 3/3 & 1/1 & - \\
            \hline
              Tiktok-2 & User$_{A}$ replies to a comment from User$_{B}$ that says `nice live!' & 0/1 & 0/3 & - & 1/1 & 0/3 & -  & 1/1 & 0/3 & - & 1/1 & 0/3 & - \\
            \hline
             \rowcolor{lightgray} Snapchat-1 & User$_{A}$ invites User$_{B}$ to real-time location sharing, and User$_{B}$ accepts the sharing request & 3/3 & 2/2 & - & 3/3 & 2/2 & -  & 3/3 & 2/2 & - & 3/3 & 2/2 & - \\
            \hline
             WeChat-1 & User$_{A}$ creates a face-to-face code, and User$_{B}$ enters the code to join the chat & 8/8 & 3/8 & - & 7/8 & 1/8 & -  & 8/8 & 3/8 & - & 8/8 & 8/8 & - \\
            \hline
             \rowcolor{lightgray} WeChat-2 & User$_{A}$ makes a video call to User$_{B}$ and User$_{C}$, and User$_{B}$ accepts the call, User$_{C}$ declines the call & 5/5 & 1/1 & 1/1 & 5/5 & 1/1 & 1/1  & 5/5 & 1/1 & 1/1 & 5/5 & 1/1 & 1/1 \\
            \hline
             Discord-1 & User$_{A}$ invites User$_{B}$ to vote an online poll, and User$_{B}$ accepts the invitation & 6/6 & 2/2 & - & 6/6 & 2/2 & -  & 6/6 & 2/2 & - & 6/6 & 2/2 & - \\
		\hline
             \rowcolor{lightgray} Discord-2 & User$_{A}$ sends an event invitation to User$_{B}$ and User$_{C}$, both accept the invitation & 4/7 & 0/3 & 0/3 & 5/7 & 2/3 & 2/3 & 5/7 & 2/3 & 2/3 & 7/7 & 3/3 & 3/3 \\
            \hline
             Douyin-1 & User$_{A}$ invites User$_{B}$ and User$_{C}$ to play together, both accept the invitation & 1/1 & 1/4 & 1/4 & 1/1 & 3/4 & 3/4  & 1/1 & 4/4 & 4/4 & 1/1 & 4/4 & 4/4 \\
            \hline
             \rowcolor{lightgray} Douyin-2 & User$_{A}$ creates a chat by face-to-face code, and User$_{B}$ enters the code to join & 6/8 & 0/8 & - & 8/8 & 0/8 & -  & 8/8 & 8/8 & - & 8/8 & 8/8 & - \\
            \hline
             Kahoot-1 & User$_{A}$ hosts a game, User$_{B}$ and User$_{C}$ joins by code \vspace{8pt} & 5/5 & 1/5 & 1/5 & 5/5 & 1/5 & 1/5  & 5/5 & 5/5 & 5/5 & 5/5 & 5/5 & 5/5 \\
		\hline
             \rowcolor{lightgray} Kahoot-2 & User$_{A}$ creates a quiz to User$_{B}$, and User$_{B}$ accepts the quiz & 2/2 & 0/1 & - & 2/2 & 0/1 & -  & 2/2 & 0/1 & - & 2/2 & 0/1 & - \\
		\hline
             GoogleMeet-1 & User$_{A}$ invites User$_{B}$ and User$_{C}$ to remote video chat, both join the chat & 5/7 & 2/4 & 2/4 & 5/7 & 2/4 & 0/4  & 5/7 & 2/4 & 2/4 & 5/7 & 4/4 & 4/4 \\
            \hline
             \rowcolor{lightgray} Gmail-1 & User$_{A}$ sends a video link to User$_{B}$ and User$_{C}$, both enter the link to join & 3/5 & 2/4 & 2/4 & 5/5 & 4/4 & 4/4  & 5/5 & 4/4 & 4/4 & 5/5 & 4/4 & 4/4 \\
            \hline
             Teams-1 & User$_{A}$ connects with User$_{B}$ by video, and User$_{B}$ accepts the connection & 2/2 & 0/2 & - & 2/2 & 2/2 & -  & 2/2 & 2/2 & - & 2/2 & 2/2 & - \\
            \hline
             \rowcolor{lightgray}  Twitch-1 & User$_{A}$ replies a message from User$_{B}$ that says `hello' \vspace{8pt} & 3/3 & 2/2 & - & 3/3 & 2/2 & -  & 3/3& 2/2 & - & 3/3 & 2/2& - \\
		\hline

             BigoLive-1 & User$_{A}$ sends a PK invitation to User$_{B}$, and User$_{B}$ accepts the invitation & 3/3 & 1/1 & - & 3/3 & 1/1 & -  & 3/3 & 1/1 & - & 3/3 & 1/1 & - \\
		\hline
             \rowcolor{lightgray} BigoLive-2 & User$_{A}$ requests for an audience connection to User$_{B}$, and User$_{B}$ accepts the request & 2/2 & 2/2 & - & 2/2 & 2/2 & -  & 2/2 & 2/2 & - & 2/2 & 2/2& - \\
		\hline
            
             BuzzCast-1 & User$_{A}$ hosts a live steam, User$_{B}$ and User$_{C}$ join the stream & 2/3 & 2/4 & 0/2 & 3/3 & 4/4 & 2/2 & 2/3 & 2/4 & 2/2 & 3/3 & 4/4 & 2/2 \\
            \hline

            \rowcolor{lightgray} Skype-1 & User$_{A}$ makes a voice call to User$_{B}$, then User$_{B}$ invites User$_{C}$ to join, and User$_{C}$ accepts the call & 5/5 & 1/1 & 1/1 & 5/5 & 1/1 & 1/1 & 5/5 & 1/1 & 1/1 & 5/5 & 1/1 & 1/1 \\
            \hline
             Skype-2 & User$_{A}$ invites User$_{B}$ to join conference meeting, then User$_{A}$ invites User$_{C}$, who accepts the invitation & 2/3 & 1/1 & 0/2 & 2/3 & 1/1 & 2/2 & 3/3 & 1/1 & 2/2 & 3/3 & 1/1 & 2/2 \\
            \hline
            
            \hline
            \multicolumn{2}{c||}{\bf{Average Success Rate}} & \multicolumn{3}{c|}{46.3\%} & \multicolumn{3}{c|}{56.1\%} & \multicolumn{3}{c|}{65.8\%} & \multicolumn{3}{c}{\textbf{82.9\%}} \\
            \hline
            \multicolumn{2}{c||}{\bf{Average Action Similarity}} & \multicolumn{3}{c|}{79.1\%} & \multicolumn{3}{c|}{89.2\%} & \multicolumn{3}{c|}{91.5\%} & \multicolumn{3}{c}{\textbf{96.8\%}} \\
            \hline
	\end{tabular}
    \vspace{0.5cm}
\end{table*}

\subsection{RQ1: Performance of \tool}
\label{sec:rq1}
\textbf{Experimental Setup.}
To answer RQ1, we evaluate the ability of our approach to accurately automate multi-user interactive features from a given task description.
While several studies~\cite{hu2018appflow,burns2022motifvln} have introduced datasets for automating or testing tasks derived from descriptions, these datasets are limited to the general features and limited to single-user settings.
Our focus on multi-user interactive features across multiple users presents a unique challenge, as there are no existing datasets tailored to this specific scenario.

To gather a collection of multi-user tasks, four authors with at least two-year backgrounds in Android development were participated.
The four annotators were then assigned to investigate the 50 popular apps on Google Play, chosen specifically for their popularity, as it often correlates with a higher likelihood of featuring multi-user interactive tasks.
We asked them to independently navigate through these apps, identify multi-user interactive tasks, and label the ground truth, which included task descriptions and the corresponding GUI execution traces.
To aid in this process, we also provided them with the apps' descriptions from Google Play, which could offer additional insights into the apps' features.
Note that we asked the annotators to set up the tasks involved to two or three users for the brevity of the experiments.
After the initial labeling, the annotators convened to meet and discuss any differences in their findings, working collaboratively until they reached a consensus on the set of multi-user interactive tasks.

To ensure the quality and practicality of the dataset, we further had two professional app developers from a large tech company to review the collected tasks.
In total, we obtained 41 multi-user interactive tasks derived from 16 apps, with an average of 2.87 actions required per user to complete each task.
Note that, all the tasks were freshly identified and labeled by human annotators specifically for this study, mitigating the potential bias for data leakage that could arise from the use of LLMs.

\textbf{Baselines.}
We set up three ablation studies as baselines to compare with our approach.
\tool is structured around two functional types of agents: user agents (i.e., Operator in Section~\ref{sec:phase2}) and supervisor agents (i.e., Coordinator in Section~\ref{sec:phase1} and Observer in Section~\ref{sec:phase3}).
As the user agents primarily focus on managing interactions with devices, we created variants of our approach that exclude the supervisor agents to assess their individual contributions.
First, we introduced a variant called \textit{Operator only}.
This variation operates solely on the Operator agents that given a multi-user interactive task, they autonomously determine the actions on the GUI screen to accomplish the task.
Second, we set up a variant, namely \textit{Operator+Coordinator}, that gains the planning capability of the Coordinator agent to deduce the number of users needed, divide the tasks into sub-tasks, and determine the sequence of interactions.
Third, to assess the importance of the Observer, we established another variant named \textit{Operator+Observer}.
In this setup, the Operator executes the task with the oversight and auditing functions that the Observer provides. 

\textbf{Metrics.}
We employ two evaluation metrics to evaluate the performance of our approach.
The first is the success rate, a commonly used metric in task automation that measures the ability of the approach to successfully complete an interactive task within an app.
A higher success rate indicates a more reliable and effective approach.
However, the success rate (either 1 or 0) is too strict without indicating the fine-grained performance of different approaches. 
Hence, we also evaluate the accuracy of the action sequences by comparing them with the ground-truth trace.
Since actions are executed in a specific order, a single accuracy metric may not be sufficient. 
Therefore, we adopt an action similarity metric from prior research~\cite{feng2022gifdroid,bernal2020translating}, which involves first computing the Longest Common Subsequence (LCS) between the inferred action sequence and the ground-truth trace.
Then, we calculate the similarity score using the formula $\frac{2 \times M}{T}$, where $M$ is the length of the LCS and $T$ is the sum of the lengths of both sequences.
This similarity score ranges from 0\% to 100\% when expressed as a percentage, with higher values indicating greater alignment with the ground truth. 
A perfect match between the generated trace and the ground truth would yield a similarity value of 100\%.


\begin{figure}
	\centering
	\includegraphics[width = 0.925\linewidth]{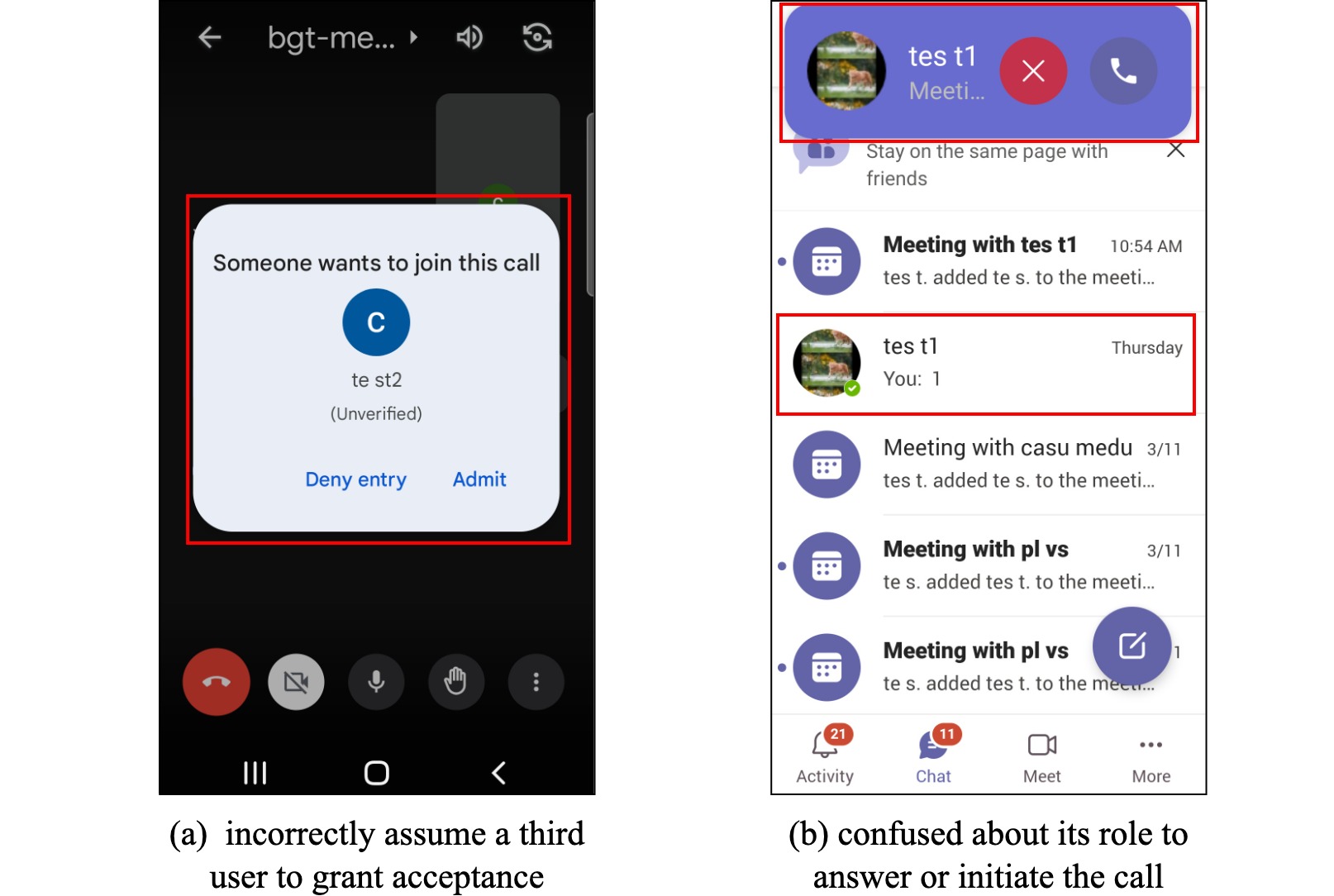}
    \vspace{5pt}
	\caption{Failure examples of ablation study of Coordinator.}
	\label{fig:rq1-1}
\end{figure}

\textbf{Results.}
Table~\ref{tab:rq1_performance_detail} presents a performance comparison between our approach and the ablation baselines. 
Due to space limitations, we present 20 representative multi-user tasks.
Note that all the results are conducted across three runs to address the potential variability of the LLMs.
Our approach achieves an average action similarity of 96.8\%, successfully completing 82.9\% of multi-user interactive tasks, which significantly surpasses the results of the ablation baselines. 
Operator only, achieves the worst performance, i.e., 46.3\% success rate and 79.1\% action similarity.
Next, we highlight the individual contributions of the Coordinator and Observer to the overall performance of \tool.

The Coordinator agent (Operator+Coordinator) enhances the success rate of task automation by 9.8\%, i.e., 56.1\% vs 46.3\% compared to Operator only.
The Coordinator's contribution is significant for two preliminary reasons.
First, it aids in determining the appropriate number of users needed for the interaction. 
For instance, in \textit{Gmail-1}, the Coordinator determines that two users are capable of achieving the task of video conference meeting, i.e., $user_A$ to initiate and control the meeting, and $user_B$ to join.
In Figure~\ref{fig:rq1-1}(a), when $user_B$ attempts to join the meeting, it requires approval from $user_A$. However, in the absence of the Coordinator, the Operator mistakenly presumes that a third user is needed to grant this acceptance.
Second, the Coordinator provides a more precise understanding of the task following segmentation, which helps to clarify the roles of the Operators involved.
For example, as shown in Figure~\ref{fig:rq1-1}(b), given the task of video communication in \textit{Teams-1}, once $user_A$ has placed the call to $user_B$ and shifts to $user_B$. However, the Operator in $user_B$ may get confused about its role by struggling to answer the call (as $user_B$) or to initiate a call (as $user_A$).

\begin{figure}
	\centering
	\includegraphics[width = 0.95\linewidth]{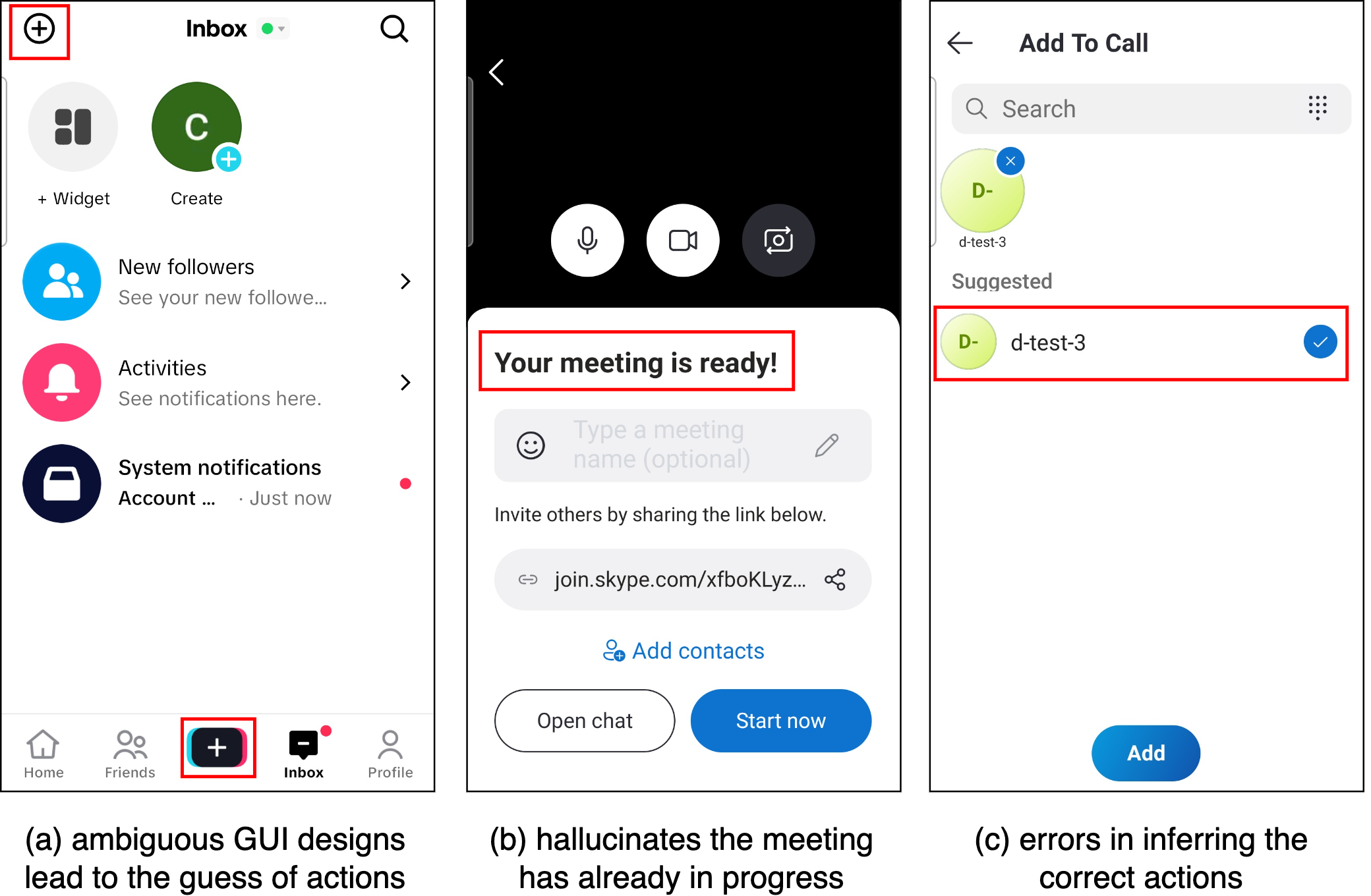}
    \vspace{5pt}
	\caption{Failure examples of ablation study of Observer.}
	\label{fig:rq1-2}
\end{figure}

Furthermore, the Observer agent (Operator+Observer) demonstrates improvements of 12.4\% in action similarity accuracy and 19.5\% in success rate, compared with Operator only.
This can be attributed to three main reasons.
First, ambiguous GUI designs can lead to confusion for the Operator.
For instance, as depicted in Figure~\ref{fig:rq1-2}(a), to invite to join a chat in \textit{Douyin-2}, the presence of two ``+'' buttons on the GUI screen may cause the approach without an Observer to mistakenly select the incorrect one, thereby struggling to break free from that erroneous exploration path.
Second, the Operator may be prone to hallucination, which can result in the generation of spurious actions based on insufficient information.
In the scenario shown in Figure~\ref{fig:rq1-2}(b), where the \textit{Skype-2} task of conference meeting, the Operator in $user_A$, unaware of the GUI state on $user_B$, might be misled to believe the text of ``Your meeting is ready'' indicates that the meeting is already in progress and switch to $user_B$, leading to a situation where $user_B$ cannot join the meeting.
Third, the Operator can become fixated on certain GUI elements, causing them to repeat actions unnecessarily.
For example, in Figure~\ref{fig:rq1-2}(c), inviting someone to join a meeting requires selecting the user and then clicking the ``Add'' button. However, the Operator powered by LLMs may incorrectly assume that simply clicking on the user's name is sufficient for the invitation.
In contrast, an Observer achieves a higher level of information, including a record of execution traces across users, resolving the issues under limited context in the Operator, such as incorrect paths, state unawareness, and repetitive actions.

\begin{figure}
	\centering
	\includegraphics[width = 0.925\linewidth]{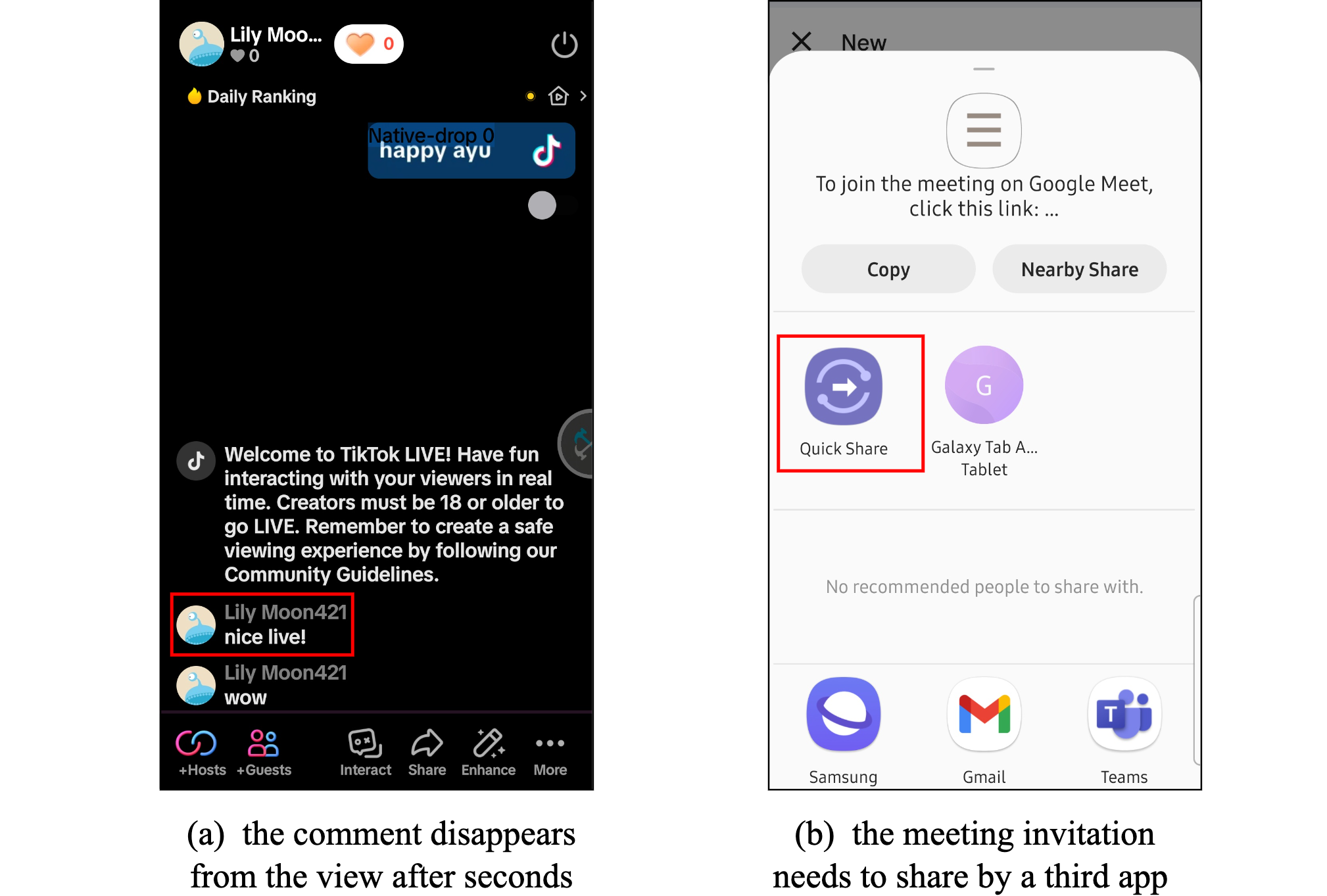}
    \vspace{5pt}
	\caption{Examples of the failure cases of our approach.}
	\label{fig:rq1-3}
\end{figure}

According to the advantages of Coordinator and Observer, our approach \tool shows a promising capability for automating multi-user interactive tasks, yet some tasks remain to fail.
The primary cause for the failures is related to timing constraints within the interactions. 
Some multi-user interactive features may require a very quick response (i.e., $<$ 2 seconds).
For instance, the \textit{Tiktok-2} task of responding to the scrolling comments in Figure~\ref{fig:rq1-3}(a); by the time the LLMs have completed its inference, the comment may have already disappeared from the view.
We are optimistic that this challenge may be mitigated by employing more advanced LLMs or by utilizing local LLMs for quicker inference, thereby reducing delays and improving the success rate of task automation.
Second, multi-user interactive features can span across multiple apps.
For instance, sharing a meeting link via an external app in \textit{GoogleMeet-1} in Figure~\ref{fig:rq1-3}(b), requires interaction with not just the meeting app but also the third-party app.
Our approach is limited by the extent to which it requires these back-and-forth interactions across different apps.

\subsection{RQ2: Comparison with State-of-the-Art}
\label{sec:rq2}
\textbf{Experimental Setup.}
To answer RQ2, we compare the performance of our approach with that of the state-of-the-art baselines. 
For this comparison, we utilize the experimental dataset collected in RQ1 (Section~\ref{sec:rq1}).

\textbf{Baselines.}
We set up eight state-of-the-art methods as baselines for comparison with our approach.
These methods include four task-driven (AdbGPT, AutoDroid, AppAgent, Mobile-agent) and four random-based (Monkey, Stoat, Sapienz, Humanoid), all commonly utilized in automated app testing.
\textit{AdbGPT}~\cite{feng2023prompting} employs the recent advancements of prompting engineering strategies such as in-context learning and chain-of-thought reasoning for automating bug tasks.
\textit{AutoDroid}~\cite{wen2024autodroid} introduces a memory injection technique to enhance LLMs with app-specific domain knowledge for task automation.
\textit{AppAgent}~\cite{zhang2023appagent} enables LLMs to autonomously interact with GUIs, learning their functionalities to achieve the tasks.
\textit{Mobile-agent}~\cite{wang2024mobile} utilizes the concept of agents to harness multimodal LLMs for task planning and navigation within mobile apps.
For a fair comparison, we added the prompt for achieving switch user action (i.e., [switch] [user] [message]) and maintained the same configuration of LLMs to avoid any evaluation bias.

\textit{Monkey}~\cite{web:monkey} is a well-known automated testing tool that generates random GUI actions to explore app features. 
\textit{Sapienz}~\cite{mao2016sapienz} introduces multi-objective search-based algorithms to automatically explore and optimize app feature testing.
\textit{Stoat}~\cite{su2017guided} leverages app state transitions to perform stochastic model-based testing on apps.
\textit{Humanoid}~\cite{li2019humanoid} utilizes a deep neural network that has been trained on real-world human interactions to simulate common feature tests.
For a fair comparison, we launched multiple devices and ran the automated testing tools on apps in hopes of triggering and covering the task.
To address potential biases of testing coverage due to time limitations, we have designated a 2-hour testing period for each tool.

\textbf{Metrics.}
To compare with the state-of-the-art baselines, we utilize the same two evaluation metrics employed in RQ1 (Section~\ref{sec:rq1}), namely success rate and action similarity. 


\renewcommand{\arraystretch}{1.05}
\begin{table}
    \small
    \tabcolsep=0.2cm
	\centering
	\caption{Performance comparison of state-of-the-art.}
	\label{tab:rq2_performance}
	\begin{tabular}{l||c|c} 
	    \hline
	   \rowcolor{darkgray} \bf{Method} & \bf{Success Rate} & \bf{ Action Similarity}\\
	    \hline
            Monkey~\cite{web:monkey} & 7.3\% & - \\
            \hline
            Sapienz~\cite{mao2016sapienz} & 9.6\% & - \\
            \hline
            Stoat~\cite{su2017guided} & 12.2\% & - \\
            \hline
            Humanoid~\cite{li2019humanoid} & 9.6\% & - \\
            \hline
            AdbGPT~\cite{feng2023prompting} & 53.7\% & 79.1\% \\
            \hline
            AutoDroid~\cite{wen2024autodroid} & 63.4\% & 85.4\% \\
            \hline
            AppAgent~\cite{zhang2023appagent} & 60.9\% & 86.8\% \\
            \hline
            Mobile-agent~\cite{wang2024mobile} & 68.3\% & 89.6\% \\
            \hline
            \hline
            \tool & \textbf{82.9\%} & \textbf{96.8\%} \\
            \hline
	\end{tabular}
\end{table}


\textbf{Results.}
Table~\ref{tab:rq2_performance} presents a performance comparison of our approach against that of the state-of-the-art baselines. 
Our approach outperforms the others across all evaluated metrics, achieving an average increase of 14.6\% in success rate and 7.2\% in action similarity over the best baseline, Mobile-agent. 
Task-driven methods consistently outperform random-based approaches, achieving average success rates of 61.6\% versus 9.7\%. 
While Mobile-agent achieves the best performance with averages of 68.3\% in success rate and 89.6\% in action similarity, it is hindered by all of the same constraints as discussed in the Section~\ref{sec:rq1} for the \tool and its ablation baselines.
Additionally, it is prone to making incorrect LLM inferences about the actions to be performed on the screen.
This discrepancy may stem from the fact that Mobile-agent's goals are slightly different from ours, being predominantly geared toward single-user scenarios than covering specific multi-user interactive features, which could lead to variations in the prompts and, consequently, a reduction in performance.

Note that since the random-based methods are not tailored to execute specific tasks, we do not calculate action similarity for them; their approach is to randomly navigate the app in hopes of accomplishing the tasks. 
Our observations indicate that Monkey, Sapienz, Stoat, and Humanoid manage to successfully complete 7.3\%, 9.6\%, 12.2\%, and 9.6\% of the multi-user interactive tasks, respectively. 
This indicates that random exploration can incidentally test some level of multi-user interactive features, particularly for simpler interactions, such as accepting a call on one device when another device initiates it. 
However, this approach falls short for more complex multi-user interactive tasks, such as joining meetings with code, which demands that the device navigates to the correct GUI screen to enter the specific code.

\subsection{RQ3: Usefulness of \tool}
\label{sec:rq3}
\textbf{Experimental Setup.}
Regression testing~\cite{thummalapenta2012automating}, a widely-known practice for task automation in software testing, involves automating previous testing tasks on new software versions to ensure that existing features remain intact and no new bugs have been introduced. 
Therefore, to answer RQ3, we assess the perceived usefulness of the \tool in facilitating regression testing, particularly for multi-user interactive features testing within mobile apps.

To carry out our study, we first collect a random sample of 18 social apps from F-Droid~\cite{web:fdroid}, in which the apps are also hosted on GitHub~\cite{web:github}.
These apps are chosen because they not only provide access to earlier versions but also offer issue-tracking platforms for bug reporting and verification.
Following the methodology described in Section~\ref{sec:rq1}, we engage four annotators and two professional developers to identify 37 multi-user interactive tasks from the apps' earlier releases.
We further confirm that these tasks remain in the next three versions.

To detect potential bugs, we automate these interactive tasks in subsequent versions of the apps. Tasks that were successfully executed in earlier versions but fail to execute in later versions may indicate the presence of bugs, such as crashes or functional errors. Based on this concept, we automate 37 tasks across three versions, resulting in 111 instances of task automation. 
We manually review instances where task automation fails to detect interactive bugs to conduct a detailed analysis of interactive bugs.

\renewcommand{\arraystretch}{0.9}
\begin{table}
    \small
    \tabcolsep=0.2cm
	\centering
	\caption{Confirmed or fixed bugs.}
	\label{tab:rq3_performance}
	\begin{tabular}{l|c|r|r} 
	    \hline
	   \rowcolor{darkgray} \bf{App Name} & \bf{Test Version} & \bf{Issue Id} & \bf{Status}\\
	    \hline
            Deku-SMS-Android & 0.32.0 & \#156 & fixed \\
		\hline
            Deku-SMS-Android & 0.32.0 & \#213 & fixed \\
            \hline
            Linphone-android & 5.1.0 & \#2039 & fixed \\
            \hline
            Linphone-android & 5.1.0 & \#2061 & fixed \\
            \hline
            Linphone-android & 5.1.0 & \#2062 & fixed \\
            \hline
            Messages & 1.1.4 & \#41 & confirmed \\
            \hline
            Phone & 5.18.1 & \#107 & fixed \\
            \hline
	\end{tabular}
\end{table}


\begin{figure}
	\centering
	\includegraphics[width = 0.90\linewidth]{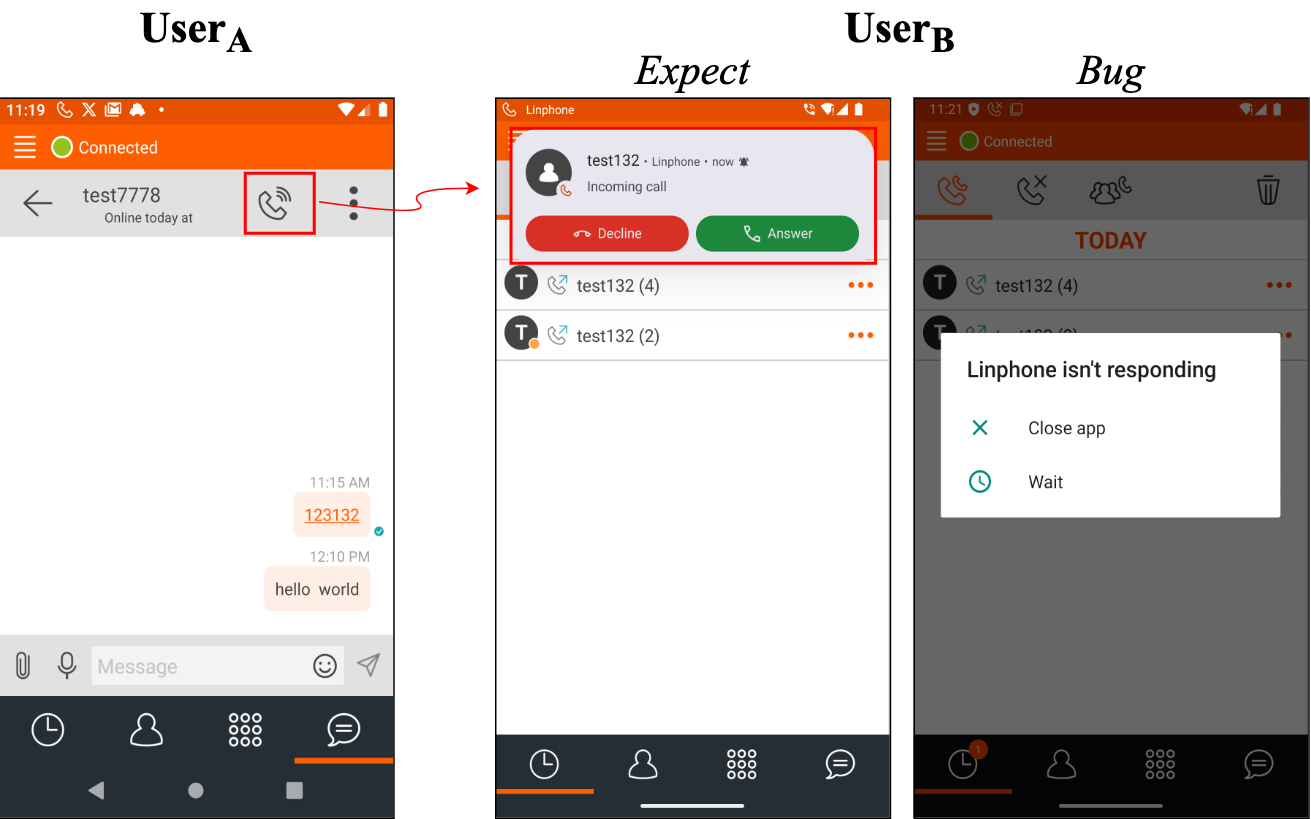}
	\caption{Illustration of interactive bug.}
	\label{fig:bug}
\end{figure}

\textbf{Results.}
Our \tool successfully completes 88.3\% of the 111 instances of task automation for regression testing.
Of the instances that failed, 11 (84.6\%) of these are identified as interactive bugs.
To further validate the effectiveness of the bugs, we cross-reference them with related issues reported in their issue-tracking repositories on GitHub, where 7 bugs have been confirmed/fixed, as shown in Table~\ref{tab:rq3_performance}.
For the remaining 4 bugs, which have not been reported earlier, we submit new issue reports under review, and as of now, none have been dismissed.
Upon further examination of the bugs, we discover that all of them require multi-user interaction to be triggered.
Figure~\ref{fig:bug} illustrates an example of a common multi-user interactive bug where one user invites another to a call, but the user does not receive it due to an app crash, hindering the automated process failure.
However, the testing capabilities of multi-user interactive features are not found in the existing automated tools, indicating the usefulness of our approach.


\section{Threats To Validity}



In our experiments to evaluate our approach, potential threats to internal validity may emerge due to the data leakage issues from LLMs.
To mitigate this bias, we recruited human annotators to label the new tasks specifically for this study as the experimental dataset to evaluate our approach.
Another potential confounding factor concerns the quality and practicality of these annotated new tasks. 
To mitigate this bias, we provided the annotators with a training session and a qualifying test before labeling. 
We further invited two professional developers from large tech companies to audit the annotated tasks for real-world feature testing.



The concern relating to external validity is the potential bias of the scalability of devices for our approach. 
While it would be ideal to perform additional experiments with a large number of devices, our existing studies involving the interaction among three devices demonstrate the potential applicability of our approach to multiple devices.
This is because the principles underlying our approach are fundamentally device-agnostic, focusing on the interaction patterns and LLM-based agent designs. 
In the future, we will systematically investigate the scalability of our approach to automate interactive feature testing across more devices.

Another factor that could affect representativeness is our use of GPT-4 as the LLM model.
\tool is designed to be model-agnostic, with a primary focus on prompt engineering. 
Recent studies~\cite{feng2025agent,feng2023prompting} usually adopt a single representative model (e.g., the state-of-the-art GPT-4) to evaluate their model-agnostic methods.
Our current experiments with GPT-based models establish a strong initial benchmark and demonstrate the potential of LLM-driven approaches for multi-user interactive feature testing. 
Evaluating the performance of \tool with other LLMs is planned as an extension toward more comprehensive experimentation in future work.

\section{Related Work}
Our research leverages a multi-agent framework to enhance the testing of multi-user interactive features within the app. Therefore, we discuss the related work, including automated app testing and multi-agent software testing. 

\subsection{Automated App Testing}

A growing body of tools has been dedicated to assisting in automated app testing, based on randomness/evolution~\cite{machiry2013dynodroid,mao2016sapienz,ye2013droidfuzzer,web:monkey}, UI understanding~\cite{chen2019gallery,xie2020uied,chen2020lost,feng2021auto,feng2022autop,xie2022psychologically,feng2022gallery,feng2023read,wang2025empirical}, UI modeling~\cite{feng2023efficiency,gu2019practical,li2017droidbot,su2017guided,feng2022gifdroid1,feng2023video2action}, and LLMs~\cite{liu2023make,wang2024feedback,feng2024mud}.
However, these tools are often constrained to test activities that are triggered by specific features.
As opposed, numerous studies focus on creating tests that target specific features, directed by manually written descriptions. 
This approach is most prominently embodied in the field of script-based record and replay techniques, such as RERAN~\cite{gomez2013reran}, Espresso~\cite{web:espresso}, WeReplay~\cite{feng2023towards}, and Sikuli~\cite{yeh2009sikuli}.
However, the reliance of these scripts on the absolute positioning of GUI elements or on brittle matching rules poses significant challenges to their adoption, as these factors can lead to a lack of robustness in automated testing.

In an effort to improve upon existing methods, many studies~\cite{wen2023empowering,feng2023prompting,feng2024enabling} aim to employ natural language descriptions that delineate the target test features.
For instance, AppFlow~\cite{hu2018appflow} generates specific tests (e.g., "add to shopping cart") based on a test library that encompasses prevalent functionalities within a given app. However, these efforts typically focus on single-user features and often overlook multi-user interactive features, which require several users to collaborate dynamically to accomplish the feature. 
This oversight can impede achieving high test coverage. 
In our work, we propose a novel approach that utilizes multi-agent LLMs to mock up several users in automating multi-user interactive tasks, thereby facilitating the app feature testing process.




\subsection{Multi-agent Software Testing}

Agents have been increasingly applied to support the automation of different software testing activities. 
One such approach relates to test management~\cite{malz2010agent,arora2018systematic}, which aims at selecting an appropriate set of test cases to be executed in every software test cycle using test unit agent and fuzzy logic.
Nevertheless, the testing of certain complex software features, such as chat functions, multi-player gaming, and interactions between users with varying levels of access, cannot be effectively conducted by a single-agent.
To address this limitation, a body of research~\cite{kumaresen2020agent,ramirez2021agent,ahlgren2020wes,ahlgren2021testing} has explored the concept of multi-agent-oriented software testing.
This involves the creation of multiple intelligent agents specifically designed to tackle the testing of these intricate features.
For instance, Tang~\cite{tang2010towards} introduced a variety of agents, each specialized in different roles equipped with hard-coded capabilities, including test design, execution, and evaluation.
Similarly, Dhavachelvan et al.~\cite{dhavachelvan2006new,dhavachelvan2005multi} have crafted several ad-hoc agents with predefined patterns to test specific features within the software.
However, these approaches rely on hard-coded rules tailored to individual software, which poses challenges for generalization across different software.

Large Language Models (LLMs) have been harnessed to facilitate a wide array of software engineering tasks.
A notable insight is that LLMs, trained on extensive datasets, can produce a compelling sense of being in the presence of a human-like interlocutor, denoted as an autonomous agent, capable of dynamically reacting to contextual cues.
Inspired by this capability, our work incorporates a multi-agent framework that integrates LLMs to address the complexities involved in testing multi-user interactive software features. 
That is, we introduce a set of autonomous agents, each with a distinct role: a Coordinator responsible for planning the interactive features, an Operator tasked with executing actions within the software environment, and an Observer assigned to track and analyze the results of these interactions.
This mirrors the collaborative aspects of human testing teams for assessing multi-user interactive features.

\section{Conclusion}
We present \tool, a novel multi-agent approach designed for automating multi-user interactive tasks for app feature testing. 
\tool innovates by deploying two functional types of multi-agents: user agents (Operator) and supervisor agents (Coordinator and Observer).
These agents are powered by the Large Language Models (LLMs), which enable autonomous actions that work collaboratively to achieve the automation of multi-user interactive tasks. 
In detail, the Coordinator directs the task; the Operator emulates the actions of interactive users; and the Observer reviews the task automation process. 
The experiments demonstrate the effectiveness and usefulness of our approach in automating multi-user interactive tasks and facilitating the regression testing process.

In the future, we plan to improve \tool in three aspects.
First, considering that multi-user interactive features are often time-sensitive, we aim to boost the efficiency of our approach by implementing open-source LLMs on the local server, hence reducing latency and improving response times. 
Second, since interactive features can extend over various applications, we will explore the potential to facilitate multi-user cross-app interactions.
Third, although our current work focuses on peer-to-peer and small-group interactions for tractability and experimental control, future research will investigate scaling to more complex social dynamics.

\begin{acks}
We sincerely appreciate the support of the Bytedance team in facilitating discussions, validating data, and conducting experiments.
This research was partially supported by OpenAI Researcher Access Program.
\end{acks}

\bibliographystyle{ACM-Reference-Format}
\bibliography{main}

\end{document}